\documentclass{article}  
\usepackage{epsf}
\usepackage{spadre2008}
\usepackage{graphicx}
\frompage{000} \topage{000}                                              %|
\def\rightmark{Strangeness in STAR}
\def\leftmark{J. Takahashi and R.Derradi de Souza}
 
\title{Strangeness production in STAR} 
\authors{
{J. Takahashi$^{1,*}$ and R. Derradi de Souza$^1$ %
}\\[2.812mm]
{\normalsize
\hspace*{-8pt}$^*$ For the STAR collaboration\\[0.2ex] 
\hspace*{-8pt}$^1$ Instituto de Fisica Gleb Wataghin, University of Campinas - UNICAMP \\ 
13083-970 Campinas, SP, Brazil \\[0.2ex] 
}}
 
\abstract{We present a summary of strangeness enhancement results
comparing data from Cu+Cu and Au+Au collisions at $\sqrt{S_{NN}}=200 GeV$ measured 
by the STAR experiment.
Relative yields in central Cu+Cu data seem to be higher than the equivalent 
sized peripheral Au+Au collision. In addition, strange particle production from these two 
systems is compared in terms of a statistical model, applying a Grand-Canonical ensemble and 
also applying a canonical correlation volume for the strange particles. Thermal fit 
results from the Grand-Canonical formalism shows little dependence on the system size but,
when considering a strange canonical ensemble, strangeness enhancement shows a strong 
dependency on the correlation volume.}

 
\begin{document}
 
\maketitle
\setcounter{page}{1}

\section{Introduction}

Strangeness production is expected to be enhanced in heavy ion collision as one of 
the possible signatures of the Quark-Gluon Plasma \cite{bib1}.
An increase of the strange quark density should result in an increase of the 
strange hyperon production thus, the measured integrated yield per participant 
should be sensitive to the density of quarks in the system.
Experimentally, strangeness enhancement has been observed in A+A collisions with 
respect to p+p collisions when comparing the normalized strange particle yields 
scaled by factors such as the number of participants in the collision, or the
number of wounded nucleons \cite{bib2,bib3}.
Thermal models with a canonical approach show that this observed 
relative strangeness enhancement can also be explained by considering a suppression
due to limitations on the phase space available for strange particles production 
in small systems \cite{bib4,bib5}. 
These models have been quite successful in describing the level of
enhancement and the observed hierarchy with the number of strange valence quarks, 
but up to now, no model has been able to describe simultaneously all features 
of the measured strangeness enhancement.
% for example the dependence of the enhancement with the collision system size and the collision energy. 
In this paper, we present the latest results from the analysis of the STAR experiment  
with respect to the strangeness enhancement including data from Au+Au and Cu+Cu 
collisions at RHIC maximum energy. In addition, we apply a statistical thermal model fit 
to study the system size dependence of thermal parameters.

\section{Strangeness Enhancement}

\begin{figure} [htb]
\centerline{\epsfxsize 2.2in \epsffile{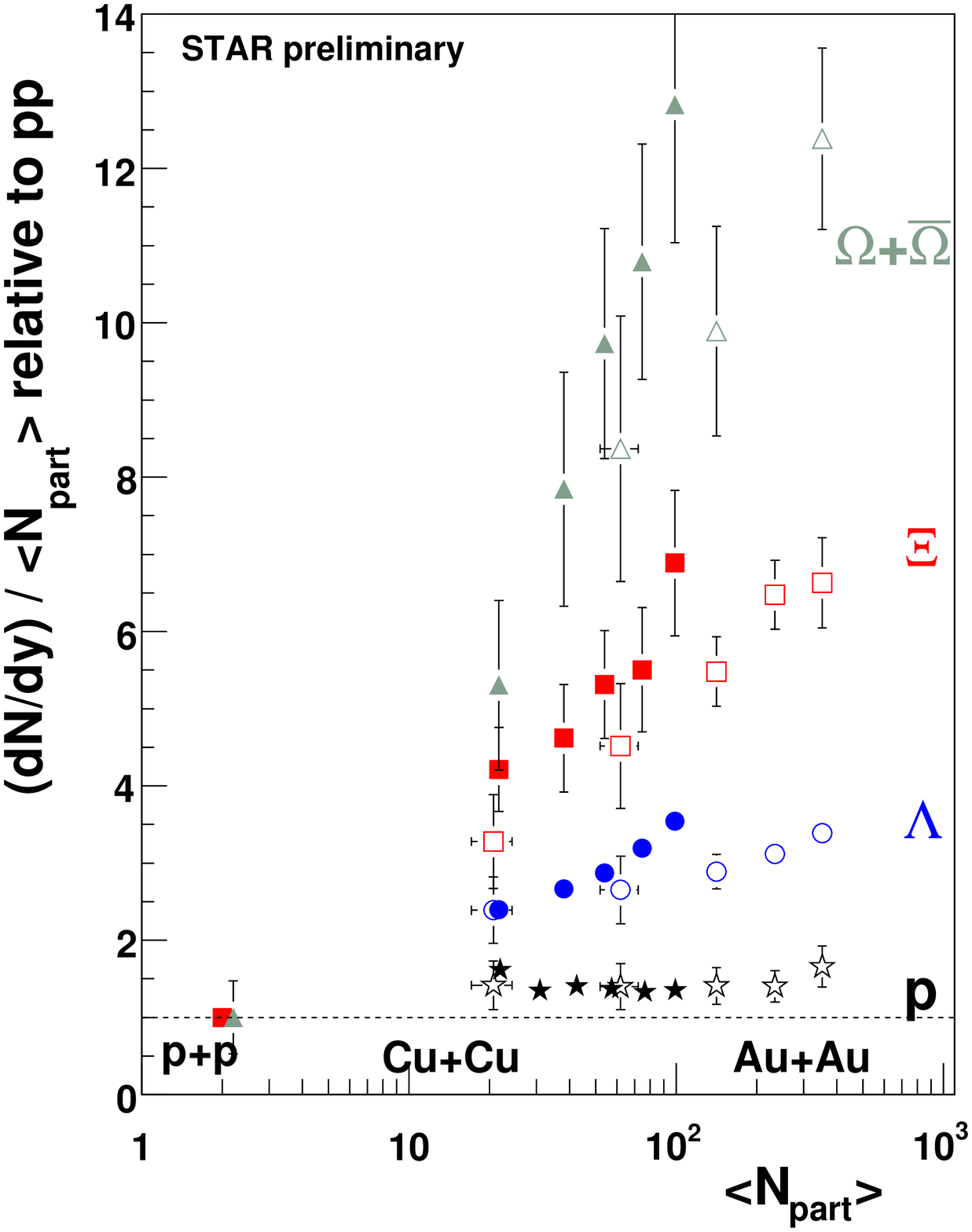} \epsfxsize 2.2in \epsffile{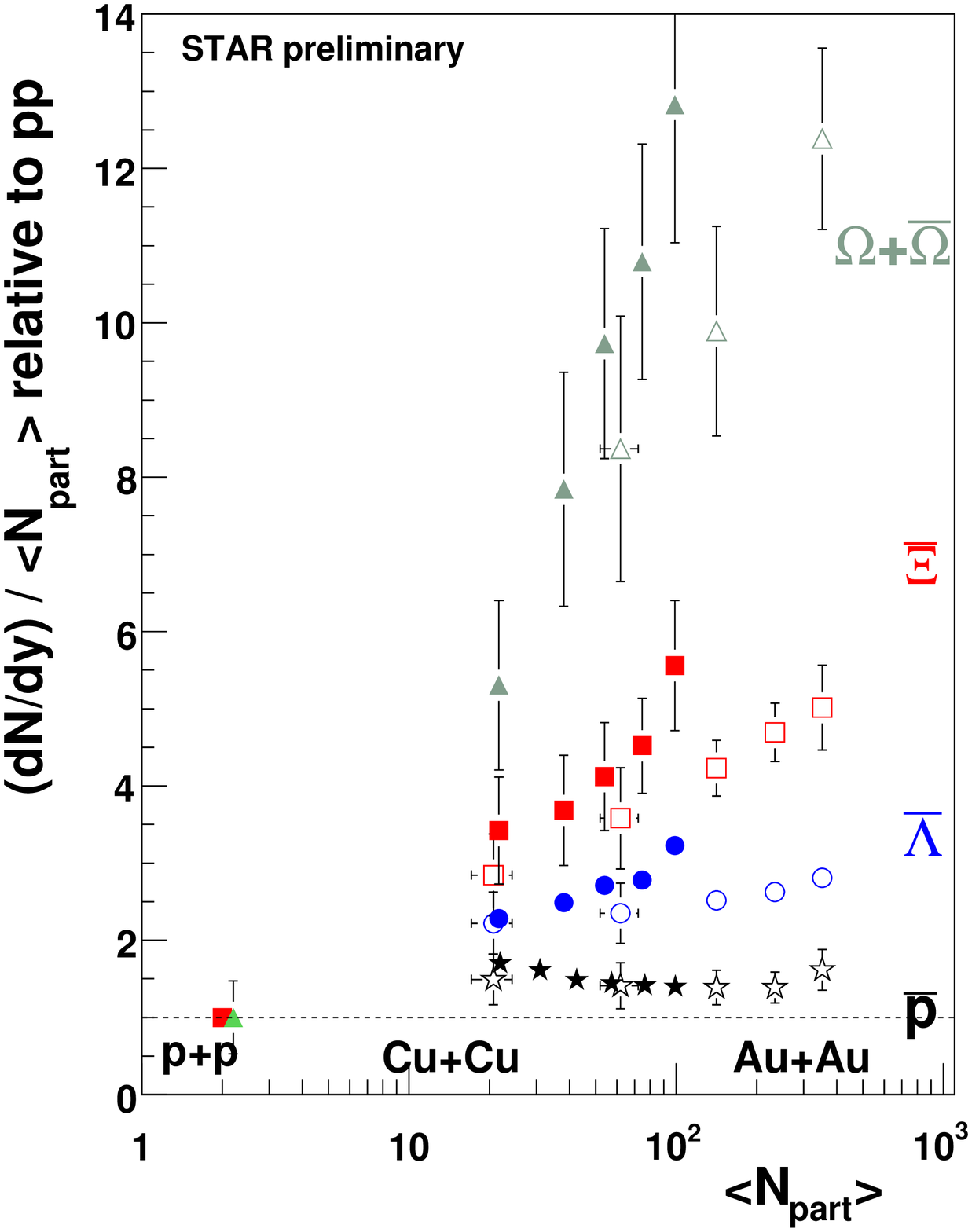}}
\caption{\label{fig1}Strangeness enhancement plots, particle yields from Au+Au and Cu+Cu 
data normalized by $<N_{part}>$ relative to yields from p+p as a function of $<N_{part}>$. 
Left plot shows the relative enhancement of $\Lambda$, $\Xi$ and $\Omega+\bar{\Omega}$, 
and the right plot shows the same data for $\bar{\Lambda}$, $\bar{\Xi}$ and $\Omega+\bar{\Omega}$.}
\end{figure}

Figure 1 left side shows the system size dependence of the strangeness enhancement 
ratio for $\Lambda$, $\Xi$ and $\Omega+\bar{\Omega}$ measured in STAR for Au+Au and 
Cu+Cu at collision energy of 200 GeV per nucleon pairs. 
Figure 1 right shows the same plot 
for $\bar{\Lambda}$, $\bar{\Xi}$ and $\Omega+\bar{\Omega}$. Particle yields were measured 
for different event centrality classes and normalized by the equivalent mean number of 
participant nucleons $<N_{part}>$ and then divided by the equivalent ratio measured in 
p+p collisions at the same energy. 
The $\Lambda$ and $\bar{\Lambda}$ yields were corrected for feed-down from $\Xi$ decays. 
In both Au+Au and Cu+Cu data, a large strangeness enhancement is observed
even in the most peripheral centrality bin. 
The data also shows a strong dependency with the system size and 
do not seem to saturate as expected from a Grand-Canonical Thermal model.  
Strangeness enhancement is higher for strange particles than for anti-particles,
which can be a result of the non zero net baryon density. The striking result 
is that the strangeness enhancement observed in central Cu+Cu data is higher 
than the peripheral Au+Au collision with equivalent $<N_{part}>$. This is an 
indication that the production mechanism of these strange particles does not 
scale with the pure geometrical parameterization of the system size.
Normalized proton yields were included in the plots of figure 1 as a reference to show 
that the relative discrepancy between the two systems is unique to the strangeness sector.
The measured proton yields in STAR are inclusive \cite{bib8}, and in this plot these 
proton yields were subtracted by the $\Lambda$ yield factorized by the decay branching 
ratio to correct for the feed-down into the proton yield. 
The normalized proton yields do not show any dependence with the system size and also
no difference between Au+Au and Cu+Cu data.

\begin{figure} [htb]
\centerline{\epsfxsize 2.5in \epsffile{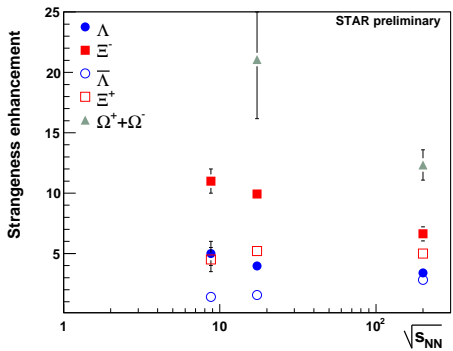}}
\caption{\label{fig2}Energy dependence of the relative strangeness enhancement for $\Lambda$,
$\bar{\Lambda}$, $\Xi$, $\bar{\Xi}$ and $\Omega+\bar{\Omega}$, observed at central collisions 
of Au+Au and Pb+Pb, compiling RHIC and SPS results. SPS results shown here are from experiment 
NA57 \cite{bib3}.}
\end{figure}

Figure 2 shows the collision energy dependence of the strangeness enhancement factor for 
the most central events for $\Lambda$, $\bar{\Lambda}$, $\Xi$, $\bar{\Xi}$, 
$\Omega+\bar{\Omega}$. SPS data for the low energy points were extracted from Pb+Pb data 
of experiment NA57 \cite{bib3}.
The observed strangeness enhancement is decreasing with collision energy 
for the $\Lambda$ and $\Xi$, as predicted by Grand-Canonical statistical model, but, 
an opposite trend is observed for the anti-particles that yields higher enhancement 
in RHIC data compared to SPS data. Also clear in this plot is the strange quark content 
hierarchy observed in the enhancement as predicted by statistical thermal models.
 
\section{Statistical thermal model results}
 
Data was analyzed using the THERMUS code \cite{bib6} and considering particle ratios 
that includes $p$ ($\bar{p}$), $\pi^{\pm}$, $K^{\pm}$, $\Lambda$ $(\bar{\Lambda})$, 
$\Xi^{\pm}$, $\Omega^{\pm}$, and $\phi$. 
Protons were corrected for the feed-down of the $\Lambda$ decay that resulted in a 
reduction of the inclusive proton yield of approximately $30\%$.
Since STAR does not yet have $\Sigma$ measurements, the feed-down contribution from this
particle was estimated using the ratio of $\Sigma/\Lambda = 0.35$ \cite{bib7}. 
Final reduction of the inclusive proton yields was on the order of $35\%$ to $45\%$. 
To estimate the error due to this feed-down correction, the relative $\Sigma$ yield 
to $\Lambda$ was varied over $50\%$. The overall effect on the final thermal parameters 
was less than $10\%$.

\begin{figure} [htb]
\centerline{\epsfxsize 2.2in \epsffile{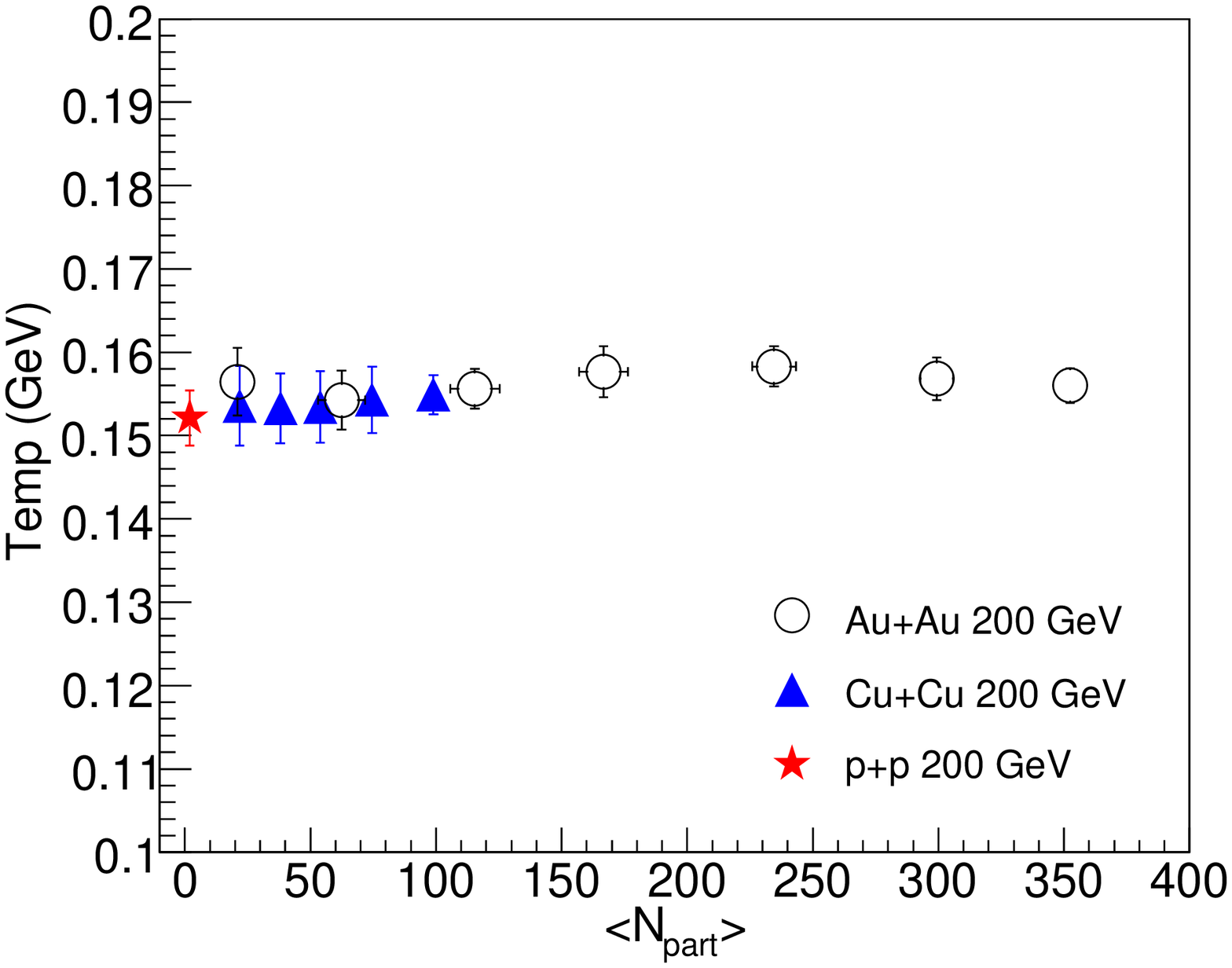} \epsfxsize 2.2in \epsffile{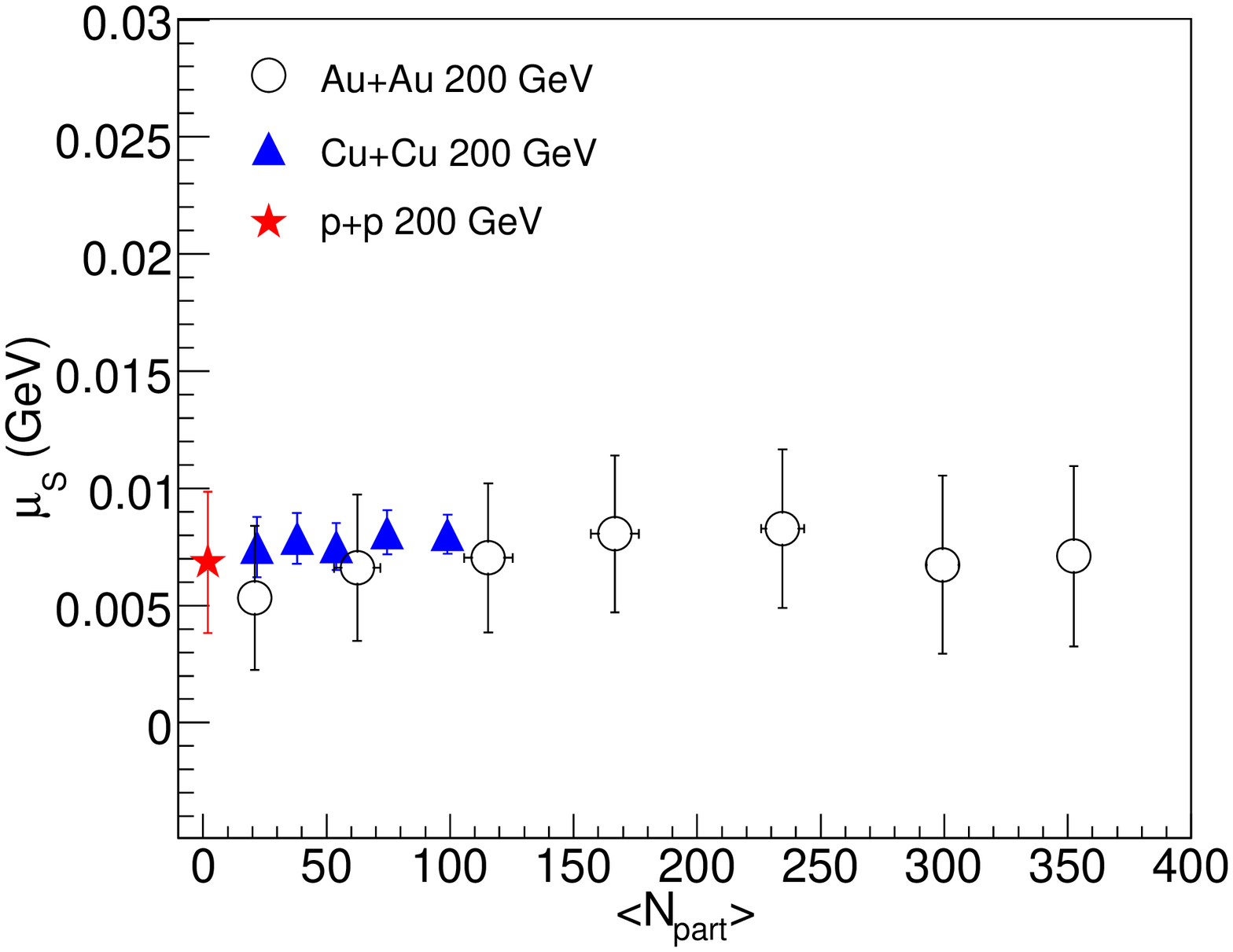}}
\caption{\label{fig3}Thermal model fit chemical freeze-out temperature $T_{ch}$, and strangeness
chemical potential $\mu_{S}$, as a function of $<N_{part}>$ for Au+Au and Cu+Cu data. 
For comparison, the same fit was applied to p+p data and included in this plot.}
\end{figure}

Figure 3 shows the chemical freeze-out temperature $T_{ch}$ and the strangeness
chemical potential $\mu_{S}$ as a function of the system size, $<N_{part}>$. 
$T_{ch}$ is around 155 MeV and seems to show no dependence on the 
system size and also no sensitivity to the colliding systems, yielding the same results 
for Au+Au and Cu+Cu data.
The baryon chemical potential $\mu_{B}$ and strangeness chemical potential $\mu_{S}$ 
also showed no difference between the fits to Au+Au and Cu+Cu data.
To study the validity of the statistical thermal fit considering a Grand Canonical 
formulation, we have applied the same fit to the p+p data. 
The chemical freeze-out temperature that results from this fit is slightly lower, around 150 MeV.

\begin{figure} [htb]
\centerline{\epsfxsize 2.5in \epsffile{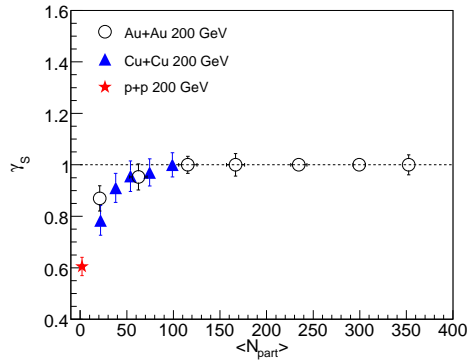}}
\caption{\label{fig4}Strangeness under-saturation parameter of the thermal model from fits to 
particle ratios of Au+Au and Cu+Cu data for the different system sizes. The result from a fit 
to the p+p data is also included.}
\end{figure}

Figure 4 shows the dependence on the system size of the strangeness under-saturation 
factor $\gamma_{S}$ obtained from the statistical thermal fit with a Grand-Canonical 
ensemble for the  Au+Au and Cu+Cu data.      
Cu+Cu data shows the same results and behavior of the peripheral Au+Au data.
The $\gamma_{S}$ reaches unity only above $<N_{part}>$ greater than approximately 100. 
Most of the Cu+Cu data is below that limit. The $\gamma_{S}$ factor 
can be interpreted as a measure of the validity of the Grand-Canonical 
formalism to describe the data in the strange sector. Under this assumption, 
from the results shown in figure 4, it is clear that only the most central bins 
of Cu+Cu collision can be well described with this approach. 
$\gamma_{S}$ from the fit to the p+p data is also shown in figure 4, and shows a 
much lower value, around 0.6, indicating that the strange particles ratios in p+p collisions 
cannot be well described with this model.
 
\begin{figure} [htb]
\centerline{\epsfxsize 2.6in \epsffile{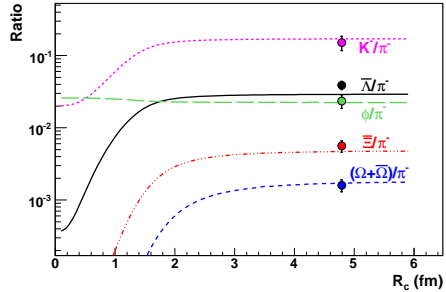}}
\caption{\label{fig5}Plot shows the strange particle to pion ratio as predicted by the 
thermal model considering a Grand-Canonical ensemble with a strange canonical system, 
as a function of the radius of the strangeness canonical size.}
\end{figure}

As an alternative to the Grand Canonical formulation (GC) where the conservation of the 
quantum numbers are ensured on average, the THERMUS code allows for a strangeness canonical 
(SC) ensemble where only the quantum numbers of the strange particles are required to 
conserve exactly. In this formalism, a correlation volume of strange particle production 
is defined as the sub-volume where the strangeness chemical equilibrium is restricted.
We applied the fit to the Au+Au 200 GeV data using this approach and obtained a 
strangeness correlation volume radius of approximately $5 fm$, for the most central event
centrality class.
This value is much higher than the value 
presented for a similar analysis on SPS data \cite{bib3} that was around $1 fm$.
To understand the effect of this correlation volume on the final strange particle yields, 
we studied the variation of the strange particle yields relative to the pions for different 
correlation volumes. Figure 5 shows the results of the model prediction in solid lines and 
the symbols represent the experimental values of the most central Au+Au 200 GeV data.
It is interesting to note that the $\phi/\pi$ ratio does not depend on the correlation volume. 
The value of the correlation volume obtained from the 
best fit to the data indicates that the system is already in the region where the strange 
particle production is saturated, and thus, consistent with a system where the strangeness 
is already equilibrated. 

\section{Conclusions}\label{concl}

Strange particle production is enhanced in heavy ion collisions compared to p+p. 
This enhancement increases with the system size and does not seem to saturate 
as expected by models considering Grand Canonical Statistical models.
When using $<N_{part}>$ as a scaling factor, Cu+Cu central collision show higher 
strangeness enhancement than peripheral Au+Au collision. 
Application of thermal model fit to Au+Au and Cu+Cu data seem to show almost no 
difference in the thermal parameters.  In addition, the thermal fit parameters 
considering a Grand-Canonical approach seem to show no dependency on the
system size except for the $\gamma_{S}$ parameter. However, when considering a 
strangeness Canonical formalism in the statistical thermal model, strangeness 
production shows strong dependence on the correlation volume. Fits to the data 
indicate that the correlation volume is large, around $5 fm$,  above the saturation point.

\section*{Acknowledgments}
We wish to thank RHIC and STAR group and the U.S. DOE Office of Science for the support 
in the participation of this conference.

%We thank the RHIC Operations Group and RCF at BNL, and the
%NERSC Center at LBNL and the resources provided by the
%Open Science Grid consortium for their support. This work 
%was supported in part by the Offices of NP and HEP within 
%the U.S. DOE Office of Science, the U.S. NSF, the Sloan 
%Foundation, the DFG Excellence Cluster EXC153 of Germany, 
%CNRS/IN2P3, RA, RPL, and EMN of France, STFC and EPSRC
%of the United Kingdom, FAPESP of Brazil, the Russian 
%Ministry of Sci. and Tech., the NNSFC, CAS, MoST, and MoE 
%of China, IRP and GA of the Czech Republic, FOM of the 
%Netherlands, DAE, DST, and CSIR of the Government of India, 
%Swiss NSF, the Polish State Committee for Scientific Research,
%Slovak Research and Development Agency, and the Korea Sci. and
% Eng. Foundation.

\vfill\eject
\end{document}